\documentclass[aps,prl,showpacs,floatfix,twocolumn,byrevtex,superscriptaddress]{revtex4-1}
\usepackage{amsmath}
\usepackage{amssymb}
\usepackage{amstext}
\usepackage{amsopn}
\usepackage{amsfonts}
\usepackage{amsxtra}
\usepackage[english]{babel}
\usepackage{graphicx}
\usepackage{float}
\usepackage{bm}
\usepackage{multirow}
\usepackage{dcolumn}
\usepackage{hyperref}

\newcommand{\icm}{\ensuremath{~\textrm{cm}^{-1}}}

\begin{document}

\title{Optical Signatures of Weyl Points in TaAs}
\author{B. Xu}
\thanks{These authors contributed equally to this work.}
\affiliation{Beijing National Laboratory for Condensed Matter Physics, Institute of Physics, Chinese Academy of Sciences, P.O. Box 603, Beijing 100190, China}
\author{Y. M. Dai}
\thanks{These authors contributed equally to this work.}
\affiliation{Center for Integrated Nanotechnologies, Los Alamos National Laboratory, Los Alamos, New Mexico 87545, USA}
\author{L. X. Zhao}
\author{K. Wang}
\author{R. Yang}
\author{W. Zhang}
\author{J. Y. Liu}
\affiliation{Beijing National Laboratory for Condensed Matter Physics, Institute of Physics, Chinese Academy of Sciences, P.O. Box 603, Beijing 100190, China}
\author{H. Xiao}
\affiliation{Beijing National Laboratory for Condensed Matter Physics, Institute of Physics, Chinese Academy of Sciences, P.O. Box 603, Beijing 100190, China}
\affiliation{Center for High Pressure Science and Technology Advanced Research, Beijing 100094, China}
\author{G. F. Chen}
\affiliation{Beijing National Laboratory for Condensed Matter Physics, Institute of Physics, Chinese Academy of Sciences, P.O. Box 603, Beijing 100190, China}
\affiliation{Collaborative Innovation Center of Quantum Matter, Beijing 100190, China}
\author{A. J. Taylor}
\affiliation{Associate Directorate for Chemistry, Life and Earth Sciences, Los Alamos National Laboratory, Los Alamos, New Mexico 87545, USA}
\author{D. A. Yarotski}
\author{R. P. Prasankumar}
\email[]{rpprasan@lanl.gov}
\affiliation{Center for Integrated Nanotechnologies, Los Alamos National Laboratory, Los Alamos, New Mexico 87545, USA}
\author{X. G. Qiu}
\email[]{xgqiu@iphy.ac.cn}
\affiliation{Beijing National Laboratory for Condensed Matter Physics, Institute of Physics, Chinese Academy of Sciences, P.O. Box 603, Beijing 100190, China}
\affiliation{Collaborative Innovation Center of Quantum Matter, Beijing 100190, China}

\date{\today}
%
%

\begin{abstract}
We present a systematic study of both the temperature and frequency dependence of the optical response in TaAs, a material that has recently been realized to host the Weyl semimetal state. Our study reveals that the optical conductivity of TaAs features a narrow Drude response alongside a conspicuous linear dependence on frequency. The width of the Drude peak decreases upon cooling, following a $T^{2}$ temperature dependence which is expected for Weyl semimetals. Two linear components with distinct slopes dominate the 5-K optical conductivity. A comparison between our experimental results and theoretical calculations suggests that the linear conductivity below $\sim$230~\icm\ is a clear signature of the Weyl points lying in very close proximity to the Fermi energy.
\end{abstract}



\maketitle

%
%
Weyl fermions represent a pair of particles with opposite chirality described by the massless solution of the Dirac equation~\cite{Weyl1929}. After a search of more than eight decades for such fermionic fundamental particles in high-energy physics, evidence for their existence remains elusive. Recently, it has been proposed that in a material with two non-degenerate bands crossing at the Fermi level in three-dimensional (3D) momentum space, the low-energy excitations can be described by Weyl equations, allowing a condensed-matter realization of Weyl fermions as quasiparticles~\cite{Wan2011,Xu2011}. The band crossing points are called Weyl points, and materials possessing such Weyl points are thus known as Weyl semimetals (WSMs).

WSMs feature peculiar band structures both in the bulk and on the surface~\cite{Wan2011,Xu2011}, leading to novel quantum phenomena that can be probed by both surface- and bulk-sensitive techniques. In the bulk of a WSM, the low-energy excitations are dominated by the Weyl points, where the bands exhibit linear dispersion in all three momentum directions, producing a 3D analogue of graphene. Weyl points can be assigned a chirality and must appear in pairs with opposite chirality. The surface state of a WSM is characterized by Fermi arcs that link the projection of the bulk Weyl points with opposite chirality in the surface Brillouin zone~\cite{Wan2011,Xu2011}. In addition, WSMs exhibit exotic magnetotransport properties due to the Adler-Bell-Jackiw chiral anomaly~\cite{Nielsen1983,Son2013,Hosur2013,Parameswaran2014}, such as large negative magnetoresistence in the presence of parallel electric ($\mathbf{E}$) and magnetic ($\mathbf{B}$) fields~\cite{Son2013,Hosur2013}.

WSMs only exist in materials where time-reversal symmetry or inversion symmetry is broken~\cite{Wan2011}. So far, several materials have been theoretically predicted to be WSM candidates, such as the pyrochlore iridates~\cite{Wan2011,Witczak-Krempa2012}, HgCr$_{2}$Se$_{4}$~\cite{Xu2011}, topological insulator heterostructures~\cite{Burkov2011,Halasz2012,Zyuzin2012}, and the solid solutions $A$Bi$_{1-x}$Sb$_{x}$Te$_{3}$ ($A$ = La or Lu)~\cite{Liu2014} and TlBi(S$_{1-x}$$R_{x}$)$_{2}$ ($R$ = Se or Te)~\cite{Singh2012} in a very narrow doping range close to the topological phase transition. However, due to the requirement of magnetic order that breaks time-reversal symmetry in large domains~\cite{Wan2011,Witczak-Krempa2012,Xu2011}, complex sample structures~\cite{Burkov2011,Halasz2012,Zyuzin2012}, or extremely precise control of the chemical composition~\cite{Liu2014,Singh2012} in these materials, none of them have been experimentally confirmed to host the WSM state.

Recently, the non-centrosymmetric and non-magnetic transition-metal monoarsenides/phosphides (TaAs, TaP, NbAs and NbP) have been predicted to be natural WSMs with 12 pairs of Weyl points~\cite{Weng2015,Huang2015NC}, triggering extensive experimental studies on these materials in search of evidence for the WSM phase~\cite{Lv2015PRX,Lv2015NP,Xu2015Sci,Xu2015NP,Yang2015NP,Xu2015arXiv,Ghimire2015,Huang2015PRX,Zhang2015a,Luo2015}. Up to date, the presence of Weyl points~\cite{Lv2015NP,Xu2015Sci,Xu2015NP,Yang2015NP,Xu2015arXiv}, the resulting Fermi arcs in the surface states~\cite{Lv2015PRX,Xu2015Sci,Xu2015NP,Yang2015NP}, and the negative magnetoresistance caused by the chiral anomaly~\cite{Huang2015PRX,Zhang2015a} have been experimentally observed in these compounds, paving the way for further investigations into WSMs.

Theoretical predictions have indicated that the WSM state also gives rise to intriguing optical responses~\cite{Burkov2011,Hosur2012,Ashby2014}. For instance, the real part of the optical conductivity $\sigma_{1}(\omega)$ at low frequencies has contributions from both free carriers (leading to a Drude response) and low-energy interband transitions in the vicinity of the Weyl points~\cite{Ashby2014}. The Drude width in a WSM is expected to vanish as $T^{2}$ upon cooling~\cite{Burkov2011,Hosur2012}, while $\sigma_{1}(\omega)$ associated with the low-energy interband transitions close to the Weyl points grows linearly with frequency (``$\omega$-linear'' conductivity). This $\omega$-linear conductivity has been observed in some Dirac-fermion quasicrystals~\cite{Timusk2013}, as well as ZrTe$_{5}$~\cite{Chen2015}, a 3D Dirac semimetal (DSM)~\cite{Li2014arXiv,Weng2014PRX,Wang2012PRB,Wang2013PRB}. Experimentally realizing the above optical properties in the transition-metal monoarsenides/phosphides will provide not only further evidence for the WSM state in these materials, but also important insights into the fundamental properties of the Weyl points.

In this Letter, we systematically investigate both the temperature and frequency dependence of the optical response in TaAs. A narrow Drude response and pronounced $\omega$-linear behaviors have been clearly observed in the optical conductivity spectra. With decreasing temperature, the Drude width diminishes as $T^{2}$, in good agreement with the expected behavior in a WSM. The low-temperature optical conductivity reveals two $\omega$-linear components with distinct slopes. By comparing these observations with first-principle calculations, we find that the $\omega$-linear conductivity below $\sim$230~\icm\ originates purely from interband transitions in the vicinity of the Weyl points that are located very close to the Fermi energy. These interband transitions contain rich information about the fundamental properties (such as the chemical potential and Fermi velocity) of the Weyl points in TaAs.

%
%

High-quality TaAs single crystals were grown with a chemical vapor transport method, as reported in Ref.~\cite{Huang2015PRX}. The frequency-dependent reflectivity $R(\omega)$ was measured at a near-normal angle of incidence on a Bruker VERTEX 80v FTIR spectrometer. In order to accurately measure the absolute $R(\omega)$ of the sample, an \emph{in situ} gold overcoating technique~\cite{Homes1993} was employed. Data from 40 to $15\,000\icm$ were collected at 11 different temperatures from 300 down to 5~K on a freshly cleaved surface in an ARS-Helitran cryostat. Since a Kramers-Kronig analysis requires a broad spectral range, $R(\omega)$ was extended to the UV range (up to $50\,000~\icm$) at room temperature with an AvaSpec-2048 $\times$ 14 fiber optic spectrometer.

%
%

\begin{figure}[tb]
\includegraphics[width=0.95\columnwidth]{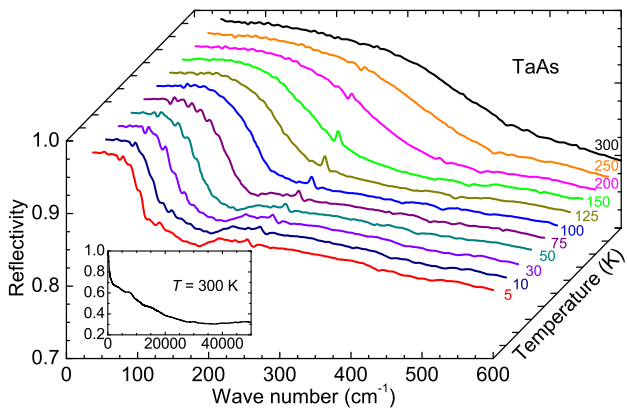}
\caption{ (color online) Waterfall plot of the far-infrared reflectivity for TaAs at different temperatures. Inset: 300-K reflectivity over a broad frequency range.}
\label{Fig1}
\end{figure}
Figure~\ref{Fig1} shows the $R(\omega)$ spectra for TaAs in the far-infrared region at 11 different temperatures. The inset displays the room-temperature $R(\omega)$ up to 50\,000~\icm. The sharp feature at $\sim$250~\icm\ is associated with an infrared-active phonon mode~\cite{Liu2015PRB}, which will not be discussed in this paper. At 300~K, $R(\omega)$ exhibits a typical metallic response characterized by a well-defined plasma edge below 500~\icm\ and a relatively high $R(\omega)$ that approaches unity at zero frequency. The low value of the plasma edge ($< 500$~\icm) generally suggests a very small carrier density~\cite{Chen2015}, consistent with the tiny volumes enclosed by the Fermi surfaces (FSs) in this material~\cite{Weng2015,Huang2015NC,Lv2015NP,Xu2015Sci,Huang2015PRX}. As the temperature decreases, the plasma edge shifts continuously towards lower frequency and becomes steeper, indicating reductions in both the carrier density and scattering rate.

The real part of the optical conductivity $\sigma_{1}(\omega)$, which provides more direct information about the charge dynamics, has been determined via a Kramers-Kronig analysis of $R(\omega)$~\cite{Dressel2002}. Below the lowest measured frequency, we used a Hagen-Rubens ($R = 1 - A\sqrt{\omega}$) form for the low-frequency extrapolation. Above the highest measured frequency, we assumed a constant reflectivity up to 12.5~eV, followed by a free-electron ($\omega^{-4}$) response.

\begin{figure}[tb]
\includegraphics[width=0.95\columnwidth]{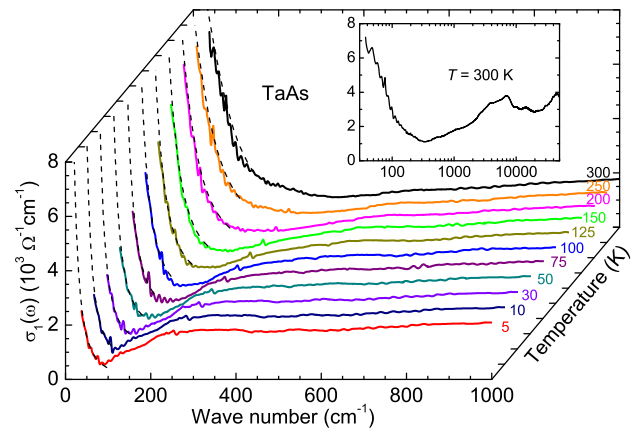}
\caption{ (color online) Waterfall plot of the temperature-dependent optical conductivity ${\sigma}_1(\omega)$ in the far-infrared region for TaAs. The dashed lines through the low-frequency data represent the fits to the Drude model. Inset: $\sigma_{1}(\omega)$ at 300~K up to 50\,000~\icm.}
\label{Fig2}
\end{figure}
Figure~\ref{Fig2} depicts the far-infrared ${\sigma}_1(\omega)$ of TaAs at all measured temperatures between 5 and 300~K, with the inset showing $\sigma_{1}(\omega)$ at 300~K up to 50\,000~\icm. At room temperature, the low-frequency $\sigma_1(\omega)$ is dominated by a narrow Drude response, indicating the metallic nature of the material. This is consistent with the reflectivity analysis. The width of the Drude peak at half maximum represents the quasiparticle scattering rate, while the area under the Drude peak (spectral weight) is proportional to the free carrier density. As the temperature is reduced, the Drude peak becomes narrower and loses its spectral weight, implying that both the quasiparticle scattering rate and carrier density drop with decreasing temperature. In addition to the Drude response, a remarkable $\omega$-linear conductivity was observed in the 300-K $\sigma_{1}(\omega)$ spectrum up to 1\,000~\icm. At 5~K, where the Drude peak becomes very small and narrow, another $\omega$-linear component with a different slope emerges in the frequency range $\sim$70--230~\icm, resulting in a noticeable kink at $\sim$230~\icm\ in the 5-K optical conductivity. Below, we will show that the $\omega$-linear component below 230~\icm\ originates purely from the low-energy interband transitions in the vicinity of the Weyl points in TaAs.

We begin with a detailed analysis of the Drude response. In order to quantify the temperature dependence of the Drude peak, we fit the low-frequency $\sigma_{1}(\omega)$ to the well-known Drude model,
%
\begin{equation}
\sigma_{1}(\omega) = \frac{2 \pi}{Z_{0}} \frac{\Omega_{p}^{2}}{\tau(\omega^{2} + \tau^{-2})},
\end{equation}
where $Z_{0} \approx 377$~$\Omega$ is the vacuum impedance; $\Omega_{p}$ and $1/\tau$ correspond to the plasma frequency and scattering rate, respectively. The fitting results are shown as dashed lines in Fig.~\ref{Fig2}, from which we notice that the low-frequency $\sigma_{1}(\omega)$ can be described quite well by the Drude model at all measured temperatures. Consequently, the temperature dependence of $\Omega_{p}$ and $1/\tau$ can be quantitatively determined from the fit.
\begin{figure}[tb]
\includegraphics[width=0.95\columnwidth]{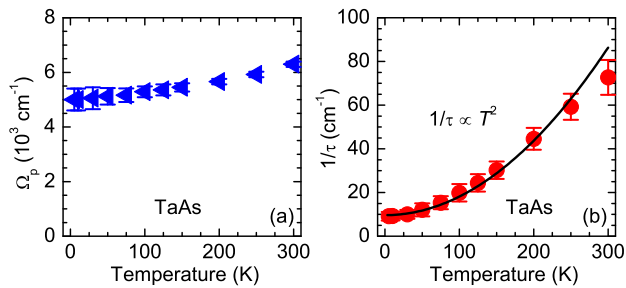}
\caption{ (color online) (a) and (b) show the temperature dependence of the plasma frequency $\Omega_{p}$ and width $1/\tau$ of the low-frequency Drude peak, respectively. The solid line in (b) is a fit to a parabolic function.}
\label{Fig3}
\end{figure}

Figure~\ref{Fig3}(a) reveals that $\Omega_{p}$ decreases slowly as the temperature is lowered, pointing to a continuous diminution of free carriers. A similar behavior has been recently reported in a DSM, ZrTe$_{5}$~\cite{Chen2015}, where the chemical potential moves towards the Dirac point as the lattice shrinks with decreasing temperature, inducing a reduction of free carriers. Such a mechanism may also apply in WSMs. Moreover, thermal excitations may also play some role in TaAs. Since the Weyl points in TaAs, which dominate the low-energy optical response, are very close to the Fermi level~\cite{Weng2015,Huang2015NC,Lv2015NP,Xu2015Sci}, the number of free carriers in proximity to the Weyl points may slightly decrease due to the reduced thermal excitations at low temperatures.

Figure~\ref{Fig3}(b) displays $1/\tau$ as a function of temperature. It is evident that the scattering rate decreases with decreasing temperature, clearly following a $T^{2}$ temperature dependence, as shown by the black solid curve through the data points. This agrees well with the expected behavior for WSMs~\cite{Burkov2011,Hosur2012}. Note that at low temperatures, $1/\tau$ becomes extremely small, in accord with the ultrahigh carrier mobility observed in transport measurements on TaAs~\cite{Huang2015PRX,Zhang2015a}.

Having systematically examined the evolution of the low-frequency Drude response with temperature, we next investigate the $\sigma_{1}(\omega)$ spectrum at different energy scales, seeking for further evidence for the existence of the Weyl points, as well as more insights into their fundamental properties. As discussed above, the Weyl points in TaAs are located in close proximity to the Fermi level~\cite{Weng2015,Huang2015NC,Lv2015NP,Xu2015Sci} and thus can be obscured by thermal excitations. To avoid this, we focus on the $\sigma_{1}(\omega)$ spectrum at the lowest measured temperature (5~K). At this temperature, the Drude peak is small and extraordinarily narrow, allowing us to access the low-energy interband transitions associated with the Weyl points.

\begin{figure}[tb]
\includegraphics[width=0.95\columnwidth]{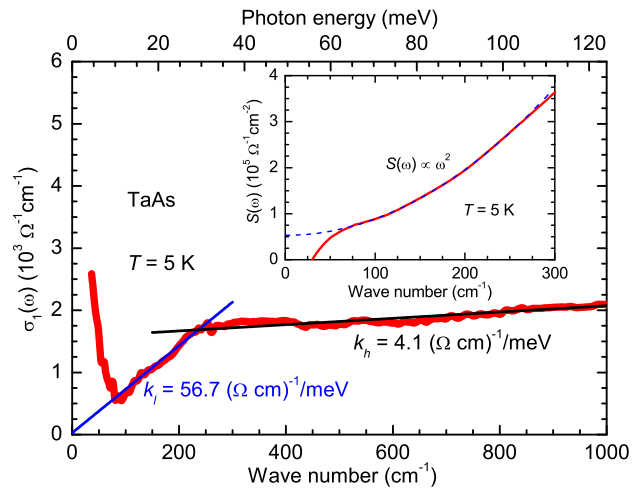}
\caption{ (color online) Optical conductivity for TaAs at 5~K. The blue and black solid lines through the data are linear guides to the eye. The inset shows the spectral weight as a function of frequency at 5~K (red solid curve), which follows an $\omega^{2}$ behavior (blue dashed line).}
\label{Fig4}
\end{figure}

Figure~\ref{Fig4} shows $\sigma_{1}(\omega)$ at 5~K up to 1\,000~\icm.
Two $\omega$-linear components with distinct slopes can be clearly identified in the conductivity spectrum, as indicated by the blue and black solid lines through the data. Both $\omega$-linear components fit well to $\sigma_{1}(\omega) = \sigma_{1}(0) + k \omega$, so that their slopes can be directly determined. The slope of the low-energy $\omega$-linear component between 70 and 230~\icm\ is $k_{l} = 56.7$~$(\Omega~\mathrm{cm})^{-1}$/meV, while the high-energy component between 230 and 1\,000~\icm\ has a slope of $k_{h} = 4.1$~$(\Omega~\mathrm{cm})^{-1}$/meV. A linearly increasing $\sigma_{1}(\omega)$ necessarily leads to a quadratic rise in the spectral weight defined as $S(\omega) = \int_{0}^{\omega}\sigma_{1}(\omega^{\prime})d\omega^{\prime}$. The inset of Fig.~\ref{Fig4} portrays the frequency dependence of $S(\omega)$ at 5~K (red solid line), which precisely follows the expected $\omega^{2}$ behavior (blue dashed line), confirming that the low-frequency optical response in TaAs is dominated by an $\omega$-linear conductivity at 5~K.

In a non-interacting electron system with two symmetric energy bands touching each other at the Fermi level, $\sigma_{1}(\omega)$ arising from interband transitions adopts a power-law frequency dependence with $\sigma_{1}(\omega) \propto (\frac{\hbar \omega}{2})^{\frac{d-2}{z}}$, where $d$ and $z$ represent the dimension of the system and the power-law of the band dispersion, respectively~\cite{Bacsi2013PRB}. This behavior has been experimentally confirmed in two-dimensional graphene ($d$ = 2 and $z$ = 1) where a frequency-independent optical conductivity was observed~\cite{Mak2008PRL}. TaAs is a 3D material ($d$ = 3). The linear conductivity is, therefore, a clear signature of linear band dispersion ($z$ = 1) which is a hallmark of WSMs.

Further information about the Weyl points in TaAs can be obtained by comparing our optical data with theoretical calculations~\cite{Hosur2012,Ashby2014}, which have demonstrated that the $\omega$-dependent optical conductivity associated with the interband transitions close to the Weyl points is given by:
%
\begin{equation}
\sigma_{1}(\omega) = \frac{N G_{0} \omega}{24 v_{F}}\Theta(\omega - 2 |\mu|)
\label{SigmaWeyl}
\end{equation}
where $N$ is the number of Weyl points, $G_{0} = 2 e^{2}/h = 7.748 \times 10^{-5}$~$\Omega^{-1}$ is the quantum conductance, $v_{F}$ is the Fermi velocity, and $\mu$ represents the chemical potential with respect to the Weyl point. This equation indicates that the $\omega$-linear conductivity arising from the Weyl points extrapolates to the origin, regardless of whether or not the Weyl points are on the Fermi surface. This agrees very well with what we observed for the low-energy $\omega$-linear component in TaAs, as shown by the blue solid line in Fig.~\ref{Fig4}, suggesting its direct link with the Weyl points.

Another deduction from Eq.~(\ref{SigmaWeyl}) is that the $\omega$-linear conductivity terminates at $\omega = 2|\mu|$, below which it falls to zero abruptly. In TaAs the low-energy $\omega$-linear component persists down to about 70~\icm\ (8.7~meV), below which it overlaps with the Drude peak. This means that $2|\mu| < 8.7$~meV, i.e. the Weyl points that give rise to the low-energy $\omega$-linear component are very close to the Fermi level. First-principle calculations~\cite{Weng2015,Huang2015NC} have predicted that TaAs possesses 12 pairs of Weyl points in total, and these Weyl points are classified into two types. Four pairs (W1) in the $k_z = 0$ plane are about 2~meV above the Fermi energy, while another eight pairs (W2) off the $k_z = 0$ plane are about 21~meV below the Fermi energy~\cite{Weng2015}. Therefore, the low-energy $\omega$-linear conductivity we observed here is associated with the low-energy interband transitions close to W1, since the interband transitions close to W2 do not turn on until $\omega > 42$~meV ($\sim$336~\icm). Moreover, interband transitions involving the trivial bands must be considered with extreme caution. Theoretical calculations have shown that in the presence of spin-orbit coupling, everywhere close to the Fermi level in the bulk Brillouin zone is gapped, with the exception of the 12 pairs of Weyl points~\cite{Weng2015,Huang2015NC}. Interband transitions involving the trivial bands do not occur below $\sim$30~meV ($\sim$240~\icm). This suggests that the $\omega$-linear conductivity below 230~\icm\ arises purely from the interband transitions in the vicinity of W1.

Equation~(\ref{SigmaWeyl}) further implies that the slope of the $\omega$-linear conductivity is directly related to $v_{F}$ through $k = \frac{N G_{0}}{24 v_{F}}$. In TaAs, there are 4 pairs of W1, i.e. $N = 8$. With the slope of the low-energy $\omega$-linear conductivity determined from the linear fit, we can calculate $v_{F} = 0.286$~eV{\AA}. Both theoretical and experimental studies~\cite{Lv2015NP,Huang2015PRX} have revealed that W1 is strongly anisotropic with Fermi velocities of 1.669, 2.835, and 0.273~eV{\AA} along $k_x$, $k_y$, and $k_z$, respectively. Optical conductivity is a momentum-averaged probe, consisting of contributions from all momentum directions. In this sense, the value of $v_{F}$ derived from our optical data falls into a reasonable range for W1.

Finally, we discuss the origin of the high-energy $\omega$-linear component (between 230 and 1\,000~\icm). The fact that it is linear in $\omega$ and turns on at a higher frequency makes it tempting to link this $\omega$-linear component with the 8 pairs of Weyl points (W2) lying 21~meV below the Fermi energy. However, as we discussed previously, on this energy scale interband transitions from the trivial bands also set in. In addition, interband transitions associated with W1 still exist. This indicates that the high-energy $\omega$-linear component is a combination of contributions from W1, W2 and the trivial bands. Nevertheless, its robust linearity implies that interband transitions close to the Weyl points may still dominate the frequency dependence of $\sigma_{1}(\omega)$, since there are 24 Weyl points in total, but only 8 trivial bands that contribute to $\sigma_{1}(\omega)$ in this frequency range. However, if we estimate $v_{F}$ from $k_{h} = 4.1$~($\Omega$ cm)$^{-1}$/meV and $N = 24$, we get $v_{F} = 11.83$~eV{\AA}. This value does not fall into a reasonable range for any of the Weyl points~\cite{Lv2015NP,Huang2015PRX}, because it is substantially larger than the values along any of the momentum directions for both W1 and W2. These facts suggest that the slope of the high-energy $\omega$-linear component is likely modified by interband transitions from the trivial bands. This may also account for its non-zero intercept.

%
%

To summarize, the optical conductivity of TaAs, a recently discovered WSM, has been measured at a variety of temperatures over a broad frequency range. We observed a narrow Drude response alongside $\omega$-linear behaviors in the optical conductivity spectra. A quadratic temperature dependence of the scattering rate was observed, in accordance with the behavior expected for a WSM. The optical conductivity at 5~K is characterized by two $\omega$-linear components with distinct slopes. The $\omega$-linear component below 230~\icm\ arises purely from the interband transitions in proximity to 4 pairs of Weyl points lying very close to the Fermi energy ($|\mu| < 4.35$~meV), providing important information about the properties of the Weyl points in TaAs. In contrast, the linear conductivity above 230~\icm\ originates from interband transitions that involve all of the Weyl points as well as the trivial bands. These experimental results thus produce the first verification of theoretical predictions for the optical response of WSMs~\cite{Burkov2011,Hosur2012,Ashby2014} in TaAs, setting the stage for further studies in this class of materials.

%
%

\begin{acknowledgments}
We acknowledge very illuminating discussions with Simin Nie, Hongming Weng, Yongkang Luo, Hu Miao, John Bowlan, Pamela Bowlan, Brian McFarland and Ricardo Lobo. Work at IOP CAS was supported by MOST (973 Projects No. 2015CB921303, 2015CB921102, 2012CB921302, and 2012CB821403), and NSFC (Grants No. 91121004, 91421304, 11374345, and U1530402). Work at LANL was performed at the Center for Integrated Nanotechnologies, a U.S. Department of Energy, Office of Basic Energy Sciences user facility, and funded by the LANL LDRD program and by the UC Office of the President under the UC Lab Fees Research Program, Grant ID No. 237789.

\end{acknowledgments}

%

\end{document}